\documentclass[letterpaper]{article}

\usepackage{aaai}
\usepackage{times}
\usepackage{helvet}
\usepackage{courier}
\usepackage{graphicx}

\usepackage[labelfont=bf]{caption}                         
\usepackage[subrefformat=parens]{subfig}
\usepackage{threeparttable}
\usepackage{amssymb}
\usepackage{ntheorem}
\usepackage{color}
\usepackage{colortbl}
\usepackage{multirow}
\usepackage{amsmath}
\usepackage{amsfonts}
\usepackage{comment}
\usepackage{bm}                                            
\usepackage{etoolbox}                                      
\usepackage{url}
\usepackage{nth}
\usepackage{natbib}
\usepackage{balance}
\usepackage{courier}                                       
\usepackage{listings}                                      
\usepackage[]{algpseudocode}                               
\algrenewcommand\textproc{\texttt}
\makeatletter\let\c@float@type\relax\makeatother
\makeatletter\let\float@addtolists\relax\makeatother
\usepackage{algorithm}

\begin{document}
%
\title{Efficient Folded Attention for 3D Medical Image Reconstruction and Segmentation}
\author{{\bf $\text{Hang Zhang}^{1,2*}$, $\text{Jinwei Zhang}^{1,2}$, $\text{Rongguang Wang}^3$, $\text{Qihao Zhang}^{1,2}$,} \\ 
 {\bf \Large $\text{Pascal Spincemaille}^2$, $\text{Thanh D. Nguyen}^2$, $\text{Yi Wang}^{1,2}$ } \\
$~^1$ Cornell University, $~^2$ Weill Cornell Medical College, $~^3$ University of Pennsylvania\\
hz459@cornell.edu
}
\maketitle

\begin{abstract}
\begin{quote}
Recently, 3D medical image reconstruction (MIR) and segmentation (MIS) based on deep neural networks have been developed with promising results, and attention mechanism has been further designed to capture global contextual information for performance enhancement.
However, the large size of 3D volume images poses a great computational challenge to traditional attention methods. 
In this paper, we propose a folded attention (FA) approach to improve the computational efficiency of traditional attention methods on 3D medical images. 
The main idea is that we apply tensor folding and unfolding operations with four permutations to build four small sub-affinity matrices to approximate the original affinity matrix.
Through four consecutive sub-attention modules of FA, each element in the feature tensor can aggregate spatial-channel information from all other elements.
Compared to traditional attention methods, with moderate improvement of accuracy, FA can substantially reduce the computational complexity and GPU memory consumption.
We demonstrate the superiority of our method on two challenging tasks for 3D MIR and MIS, which are quantitative susceptibility mapping and multiple sclerosis lesion segmentation.
\end{quote}
\end{abstract}

\section{Introduction}

\label{sec:introduction}

Recent deep convolutional neural networks (CNNs) are driving advances in various computer vision tasks.
These tasks include high-level image recognition~\cite{krizhevsky2012imagenet, simonyan2015vgg}, object detection~\cite{ren2015faster, law2018cornernet}, and semantic segmentation~\cite{fu2019dual}. 
CNN also significantly improves the performance of several low-level tasks such as super resolution~\cite{dong2014learning} and image denoising~\cite{yang2017deep}, where full functional mapping between source and target images is required.
Besides the breakthrough of natural image processing, medical image processing also benefits from CNN in various aspects.
CNN based methods surpass traditional methods and achieve the near-radiologist-level performance on MRI brain tumor segmentation~\cite{myronenko20183d}, MRI multiple sclerosis segmentation~\cite{zhang2019rsanet}, and CT Pulmonary Nodule Detection~\cite{setio2016pulmonary}, etc. 
For full functional mapping task, CNNs~\cite{yoon2018quantitative, zhang2020extending} also outperform traditional optimization-based 3D MRI image reconstruction~\cite{liu2012morphology} that requires hand-crafted regularizers or priors.

\begin{figure}[!t]
\centering 
{
\subfloat{\includegraphics[width=0.24\textwidth]{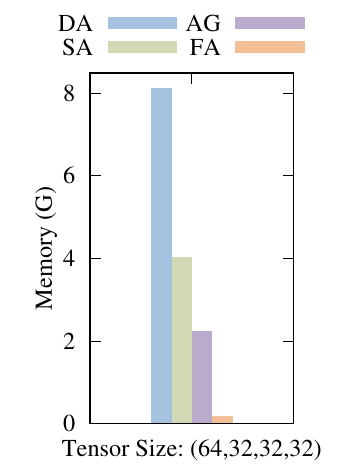}}
\subfloat{\includegraphics[width=0.24\textwidth]{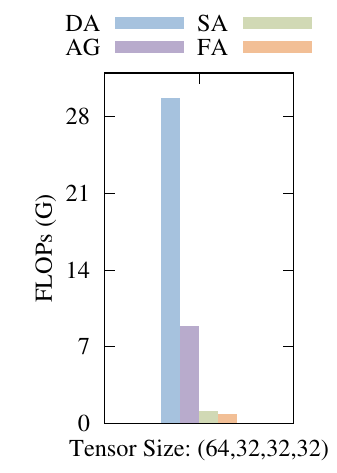}}
} 
\caption{Computational comparison of GPU memory and floating point number operations per second (FLOPs) between four different attention approaches: DA~\cite{fu2019dual}, AG~\cite{oktay2018attention}, SA~\cite{wang2018non}, and our FA. 
We get all the numbers from a machine with a single Titan Xp GPU. 
We test each module using a input feature tensor with size ($64\times32\times32\times32$). 
It can be seen that our FA module can substantially reduce computational cost compared to DA, AG and SA modules ($97.9\%$, $92.5\%$ and $95.8\%$ of GPU memory reduction, and $88.9\%$, $63.0\%$ and $25.6\%$ of FLOPs reduction).
}
\label{fig:memory_computation}
\vskip -0.1in
\end{figure}

These CNN models benefit from capturing contextual information that is essential for many computer vision tasks.
Traditional models~\cite{krizhevsky2012imagenet, simonyan2015vgg, he2016deep, huang2017densely} stack many layers of convolutional operations to capture the global contextual dependency.
However, this stacking procedure has three major drawbacks: 1) Too many convolution layers will introduce redundant network parameters that causes unnecessary memory usage and computational overhead, and makes it prone to overfitting~\cite{simonyan2015vgg, peng2017large}; 2) Network optimization becomes increasingly difficult as the network depth increases~\cite{he2016deep, huang2017densely}; 3) Information propagation among elements with large spatial distances in the feature tensor can be inefficient due to the issue of vanishing gradients~\cite{he2016deep} and saturated activations~\cite{ioffe2015batch}.

The recent attention methods have shed light to the above issues.
Self-attention methods~\cite{wang2018non, zhang2019rsanet, huang2019ccnet, fu2019dual} aim at capturing long-range dependencies by aggregating contextual information of each pixel from all other pixels in the feature map (pixel here indicates the feature vector of a pixel). 
Another stream of attention methods~\cite{wang2017residual, hu2018squeeze, oktay2018attention} focus on creating a mask that can implicitly assist CNN to pay more attention to salient areas.
Most of these attention methods operate either on spatial dimensions~\cite{zhang2019rsanet, wang2018non, oktay2018attention, wang2017residual}, or solely on the channel dimension~\cite{hu2018squeeze}, which reduces the performance of feature aggregation.
Besides, unlike natural images, processing 3D medical images using CNNs usually demands high GPU memory usage, and most of these methods are not satisfactory due to the computation of huge attention maps.
We argue that a unified attention approach that considers both the spatial-channel dependency and the efficiency of computation is of great practical value for modern 3D Medical image tasks.  
In this paper, we present our folded attention (FA) approach, effective and yet efficient, for modeling the global contextual information with negligible computational cost (see Fig.~\ref{fig:memory_computation} and Fig.~\ref{fig:memory_computation_overtime}).

\begin{figure}[!t]
\centering
\includegraphics[width=0.48\textwidth,height=0.2553\textwidth]{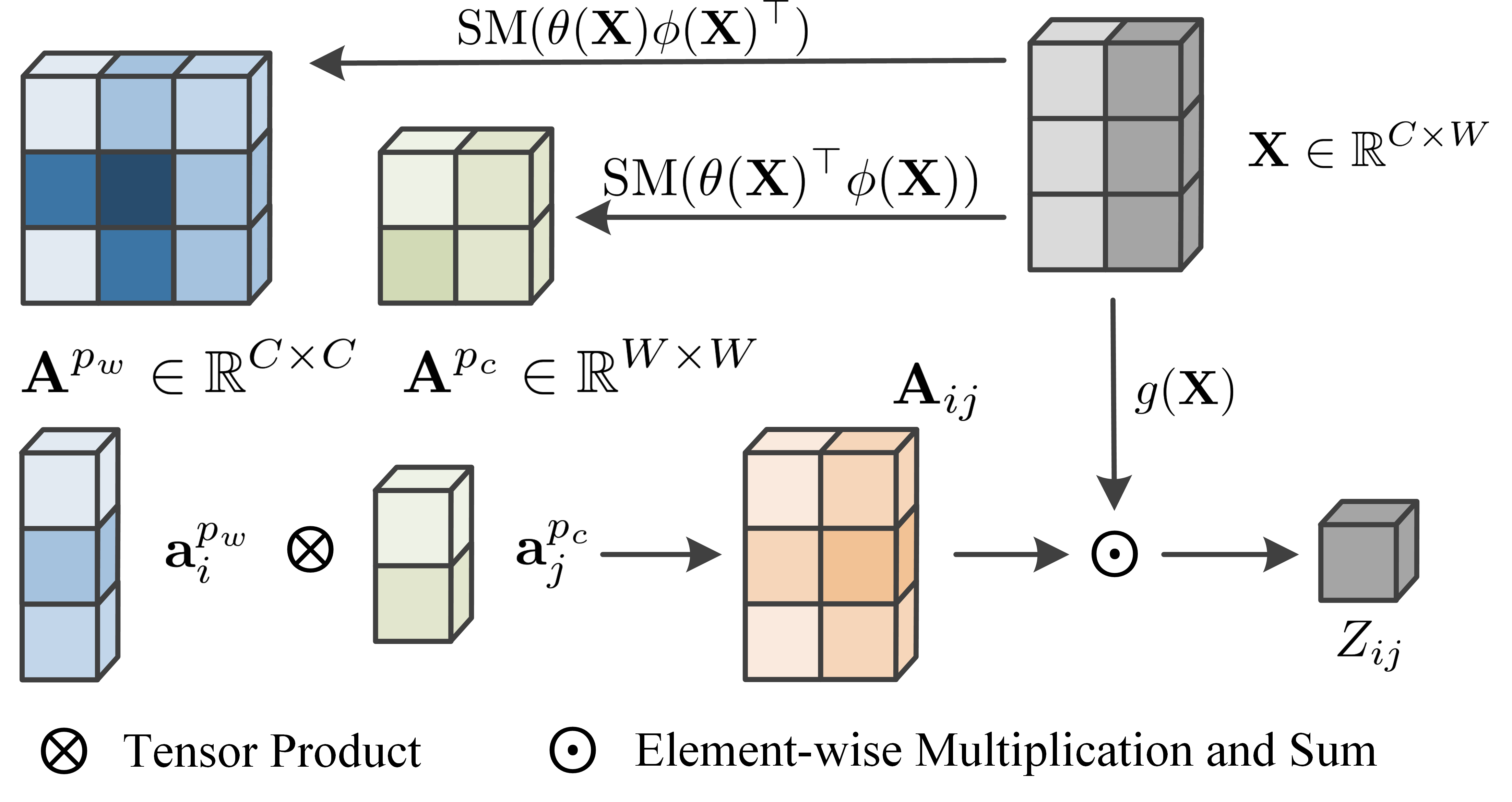}
\caption{Example illustration of the regularization of FA.}
\label{fig:fa_decomposition}
\vskip -0.1in
\end{figure}

\begin{figure*}[t]
	\centering
	\includegraphics[width=0.98\textwidth,height=0.1779\textwidth]{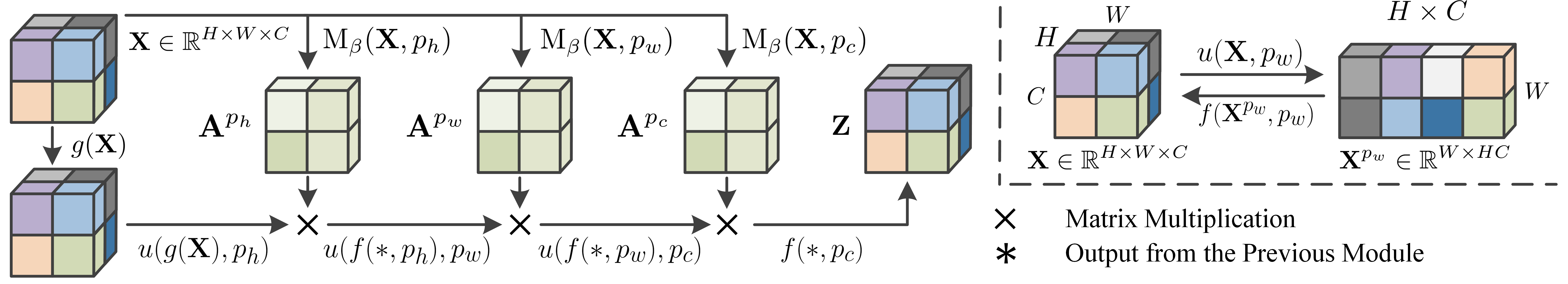}
	\caption{The left panel is the overall FA pipeline and the upper right panel is the visualization of fold and unfold operations.
		For simplicity and visualization-friendly, we use a 3D input tensor to illustrate, but 4D or higher dimensional tensors can be easily extended. 
		In the left panel, $P_h=(0,1,2)$, $P_w=(1,0,2)$ and $P_c=(3,0,1)$; we first compute three sub-affinity matrices $\mathbf{A}^{p_h}$, $\mathbf{A}^{p_w}$, and $\mathbf{A}^{p_c}$; With these sub-affinity matrices, we then use three consecutive unfolding-and-folding steps to perform the feature aggregation and get output $\mathbf{Z}$.
		In the right panel, we show how function $f$ and $u$ works; each element of the tensor is marked with a different color, and the color remains its position after folding or unfolding operations. (best view in color)
	}
	\label{fig:fa_concept}
	\vskip -0.1in
\end{figure*}

Our FA approach can be considered as the generalization of original self-attention (SA) mechanism~\cite{wang2018non}.
The original SA ignores channel-wise dependency and only aggregates information from spatial domain, while in our FA, each element in the output feature tensor is a weighted sum of all elements in the input feature tensor (a pixel is denoted as a vector with multiple elements).
Channel information does help the network learn better semantic information~\cite{fu2019dual, hu2018squeeze}, but directly applying SA to incorporate spatial and channel information will cause unacceptable GPU memory usage (more details in the methodology section).
Though DA network~\cite{fu2019dual} combines spatial and channel attention by element-wise sum operation, it still suffers from the heavy computational cost. (see Fig.~\ref{fig:memory_computation})
FA module resolves the issue by introducing tensor folding and unfolding operations, where the input feature tensor will be broadcast and unfolded to compute four sub-affinity matrices that can approximate the function of original affinity matrix with cascaded aggregation. (see Fig.~\ref{fig:fa_concept})

Through the approximation, FA can also be considered as the regularization of the SA mechanism.
For simplicity and to be visually interpretable, we use a 1D image with a 2D feature tensor to illustrate the concept and equations will be presented later.
As shown in Fig.~\ref{fig:fa_decomposition}, we use two smaller sub-affinity matrices $\mathbf{A}^{p_w}$ and $\mathbf{A}^{p_c}$ to replace the original element-to-element affinity matrix $\mathbf{A} \in \mathbb{R}^{CW\times CW}$.
Let $\mathbf{Z}$ denotes the matrix obtained after FA operation to $\mathbf{X}$ and  then $\mathbf{Z}$ can be constructed as follows:
\begin{align}
\mathbf{A}_{ij} & = \mathbf{a}^{p_w}_i \otimes \mathbf{a}^{p_c}_j, \\
Z_{ij} & = \mathbf{A}_{ij} \odot g(\mathbf{X}), 
\end{align}
where $\otimes$ is the tensor product, $\odot$ is the element-wise multiplication and sum, $\mathbf{A}_{ij}$ is the affinity matrix of element $X_{ij}$ ($A_{ij,pq}$ denotes the entry at $pq$ of matrix$\mathbf{A}_{ij}$, and is also the affinity between element $\mathbf{X}_{ij}$ and $\mathbf{X}_{pq}$ ), and $\mathbf{a}^{p_w}_i$ and $\mathbf{a}^{p_c}_j$ denote the transpose of $i_{th}$ and $j_{th}$ row of matrix $\mathbf{A}^{p_w}$ and $\mathbf{A}^{p_c}$ respectively.
It is obvious that $\mathbf{A}_{ij}$ is a rank-one matrix thus imposing regularization on the dense affinity matrix.

Our FA approach can be applied to many other 3D image analysis tasks as it is efficiency and simple, and in this paper, we demonstrate its efficiency and effectiveness on two challenging tasks in 3D Medical images. 
One task is quantitative susceptibility mapping (QSM)~\cite{de2010quantitative, wang2015quantitative}, a low-level full image mapping task.
This reconstruction problem is challenging as it needs to solve an ill-conditioned dipole inversion problem, where training data can only be obtained with COSMOS~\cite{liu2009calculation} and only very limited data samples are available~\cite{yoon2018quantitative,zhang2020fidelity}.
Another task is multiple sclerosis (MS) lesion segmentation, a high-level image segmentation task.
Unlike tumor or other organ segmentation problems, MS lesion segmentation is more difficult as lesions vary enormously in terms of size, shape, location, and conspicuity. 

\subsection{Related Works}
The attention concept is first introduced in neural machine translation~\cite{bahdanau2014neural, luong2015effective} to improve the performance of recurrent neural networks (RNN) by capturing dependencies between long-range words in a sentence.
Later, RNN is entirely replaced with self-attention operations by transformer~\cite{vaswani2017attention}.
Further, attention mechanism has then been widely adopted in vision tasks, such as image recognition~\cite{wang2017residual, hu2018squeeze}, and image segmentation~\cite{zhang2019rsanet, fu2019dual}.
In general, most of these attention methods can be divided into two types:  mask-based attention (MA) that learns a salience feature map and self-attention (SA) that learns feature aggregation. 
MA methods usually generate a mask that emphasizes the importance or saliency on a portion of the feature tensor, either spatial-wise~\cite{wang2017residual, oktay2018attention} or channel-wise~\cite{hu2018squeeze}.
Though AG-Net~\cite{oktay2018attention} improves by using grid-based gating scheme, MA methods is not suitable for image-to-image functional mapping tasks as any pixel matters and salient area is unnecessary. 
SA methods produce a function that pass through a feature map without any modification of the input size, and features either from spatial locations~\cite{zhang2019rsanet, wang2018non} or channel maps~\cite{fu2019dual} are aggregated during the pass, where each element is replaced with a weighted sum of features from some of other elements.  
SA methods raise memory issue in 3D medical images as it needs to compute huge attention maps (See DA and SA in Fig.~\ref{fig:memory_computation}).
Though RSA-Net~\cite{zhang2019rsanet} solves memory problem by iterative feature aggregation, it ignores the channel information aggregation.

\subsection{Contributions}
In this paper, we propose a novel FA approach that can efficiently capture global contextual dependencies with negligible computational cost.
We exploit the superiority of FA in QSM reconstruction and MS lesion segmentation tasks, and the contributions of FA can be summarized as follows:
\begin{itemize}
    \item We propose a folded attention approach that can improve the performance of general 3D medical image tasks by global contextual information aggregation, and our method can tremendously reduce the computational cost of GPU memory (at least $95.8\%$) and FLOPs (at least $25.6\%$) compared to the most existing attention approaches.
    \item Extensive experimental results from both high-level segmentation task and low-level image mapping task on 3D medical images show the effectiveness and the efficiency of our method. By insertion of our FA module, with negligible cost, we outperform all other attention methods, and improve the baseline Dice metric of MS lesion segmentation by $3\%$ and the baseline RMSE metric of QSM reconstruction by $3\%$.
\end{itemize}

\section{Methodology}

\label{sec:methodology}

In this section, we will present details of the proposed folded attention (FA) approach.
We will first review traditional SA mechanism and its simple generalization to channel dimension.
We then illustrate how our FA approach can generalize and regularize the SA mechanism.
Complexity analysis of memory and computational cost on FA will be discussed.


\subsection{Self Attention Mechanism}

In this paper, we adopt a widely used instantiation of SA as the rest shares similar performance~\cite{wang2018non}.
The adopted embedded Gaussian SA can be described as follows:
\begin{align}
    \mathbf{A} & = \text{SM}(\theta(\mathbf{X})\phi(\mathbf{X})^{\top}), \\
    \mathbf{Z} & = \mathbf{A}g(\mathbf{X}), 
\end{align}
where $\text{SM}$ is the Softmax function along each matrix row, $\mathbf{X} \in \mathbb{R}^{N\times C}$ is the input feature tensor, $\mathbf{A} \in \mathbb{R}^{N\times N}$ is the affinity matrix, $\mathbf{Z} \in \mathbb{R}^{N\times C}$ is the output feature tensor of SA, and $N=HWD$ is the number of pixels in the image.
Function $\theta$ and $\phi$ are single-layer perceptrons that can linearly transform features of $\mathbf{X}$ to facilitate the computation of affinity matrix.
The inner product between $\theta$ and $\phi$ computes the pixel-to-pixel affinity.
Function $g$ is also a single-layer perceptron that can help the network to learn a better feature embedding. 

\subsection{Generalization of SA to Channel Dimension}

DA-Net~\cite{fu2019dual} uses the element-wise sum of the outputs of spatial attention and channel attention to approximate spatial-channel attention. 
However, separate operations on spatial and channel dimensions are prone to be sub-optimal.
One natural idea to generalize the original SA is to replace $\mathbf{X} \in \mathbb{R}^{N\times C}$ as $\hat{\mathbf{X}} \in \mathbb{R}^{NC \times 1}$, where element-to-element instead of pixel-to-pixel affinity matrix can be obtained by $\hat{\mathbf{A}} \in \mathbb{R}^{NC\times NC}$.
Unfortunately, the matrix $\hat{\mathbf{A}}$ is too huge for modern commercial GPU to process. (According to our experiments, $\hat{\mathbf{A}}$ may consume several hundred Gigabytes memory on our 3D image tasks.)
It is obvious that direct computation of such huge matrix $\hat{\mathbf{A}}$ is not realistic, thus we propose our FA to ease the problem.

\subsection{Folded Attention (FA)}

We here propose our FA approach that can ease the problems in current attention methods: 1) FA considers spatial-channel attention in a single module; 2) FA consumes negligible computational resources. 
We will introduce the folding and unfolding operations, followed by sub-affinity matrix computation and feature aggregation.

\subsubsection{Fold and Unfold Operations}
Let $\mathbf{X} \in \mathbb{R}^{H\times W \times D \times C}$, where $H,W,D$ are sizes of three spatial dimensions of the feature tensor $\mathbf{X}$ and $C$ is the number of channels. 
We define an unfold function $u$, where $u(\mathbf{X}, p) = \mathbf{X}^p$, and $P$ is a permutation that indicates how to unfold the tensor.
Here we use an example to illustrate the function $u$.
Let $p=(1,0,2,3)$, we can get $u(\mathbf{X}, p) = \mathbf{X}^p \in \mathbb{R}^{W \times HDC}$, where $u$ first permutes the four dimensions of $\mathbf{X}$ according to $p$, and then $f$ unfolds the last three dimensions into one dimension, resulting in a 2D matrix $\mathbf{X}^p$.
Also, we define a function $f$ as the inverse operation of $u$, where $f(\mathbf{X}^p, p) = \mathbf{X}$.
For simplicity, we further set four permutation vectors as $p_h=(0,1,2,3)$, $p_w=(1,0,2,3)$, $p_d=(2,0,1,3)$, and $p_c=(3,0,1,2)$. 

\subsubsection{Sub-Affinity Matrix}

The generalization of SA with channel attention requires the computation of a huge affinity matrix $\hat{\mathbf{A}}$, which suffers from heavy memory cost.
In our proposed FA approach, we use four sub-affinity matrices to replace the huge one.
We denote the four matrices as $\mathbf{A}^{p_h}$, $\mathbf{A}^{p_w}$, $\mathbf{A}^{p_d}$, and $\mathbf{A}^{p_c}$, where $p_h,...,p_c$ are the permutation vectors defined in the last section.
The sub-affinity matrix can be computed as follows:
\begin{equation}
    \mathbf{A}^p = \text{SM}(u(\theta(\mathbf{X}), p)u(\phi(\mathbf{X}), p)^{\top}).
    \label{eq:sub_affinity_matrix}
\end{equation}
The size of each sub-affinity matrix $\mathbf{A}^p$ is much smaller than the original affinity matrix $\hat{\mathbf{A}}$.
Even the sum of the sizes of all four sub-affinity matrices is several orders of magnitude smaller than $\hat{\mathbf{A}}$. (see more details in complexity analysis)

\begin{figure}[!t]
\centering 
{
\subfloat[FLOPs over input size]{\includegraphics[width=0.4\textwidth]{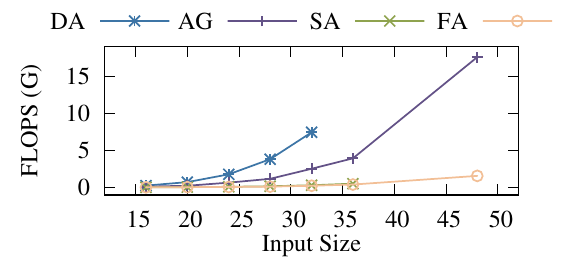} 
\label{fig:memory_computation_overtime_a}} \\
\subfloat[Memory over input size]{\includegraphics[width=0.4\textwidth]{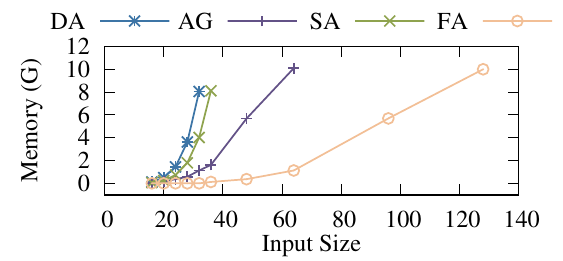}
\label{fig:memory_computation_overtime_b}}
} 
\caption{The $x$-axis is the size of the input 4D feature tensor (all four spatial-channel dimensions have equal sizes).
The $y$-axis represents the computational cost measured by FLOPs and GPU memory consumption for (a) and (b) respectively.
All numbers are obtained with a Titan XP GPU.
}
\label{fig:memory_computation_overtime}
\vskip -0.1in
\end{figure}

\subsubsection{Feature Aggregation}

The next step after obtaining affinity matrix is to aggregate features from the original feature tensor $\mathbf{X}$.
Suppose we have obtained a sub-affinity matrix $\mathbf{A}^{p}$ from Eq.~\eqref{eq:sub_affinity_matrix}, the feature aggregation based on the sub-affinity matrix can be described as follows:
\begin{equation}
    \mathbf{Z} = f(\mathbf{A}^{p}u(g(\mathbf{X}), p),p).
    \label{eq:fea_aggregate}
\end{equation}
For simplicity, we denote Eq.~\eqref{eq:fea_aggregate} as: $\mathbf{Z} = \text{U}_\gamma(g(\mathbf{X}), p)$, where $\gamma$ represents the parameters of the function $g$.
Also, Eq.~\eqref{eq:sub_affinity_matrix} can be simplified as $\text{M}_{\beta}(\mathbf{X}, p)$, where $\beta$ denotes the parameters of function $\theta$ and $\phi$.  
We can then get our four sub-affinity matrices by $\mathbf{A}^{p_h} = \text{M}_{\beta}(\mathbf{X}, p_h)$, $\mathbf{A}^{p_w} = \text{M}_{\beta}(\mathbf{X}, p_w)$, $\mathbf{A}^{p_d} = \text{M}_{\beta}(\mathbf{X}, p_d)$, and $\mathbf{A}^{p_c} = \text{M}_{\beta}(\mathbf{X}, p_c)$.
Now our proposed FA operation is derived as follows:
\begin{equation}
    \mathbf{Z} = \text{U}_{\gamma}(\text{U}_{\gamma}(\text{U}_{\gamma}(\text{U}_{\gamma}(\mathbf{X}, p_h), p_w), p_d), p_c)
    \label{eq:full_fa}
\end{equation}

\begin{figure*}[!ht]
\centering 
{
\subfloat[T1]{\includegraphics[width=0.17\textwidth]{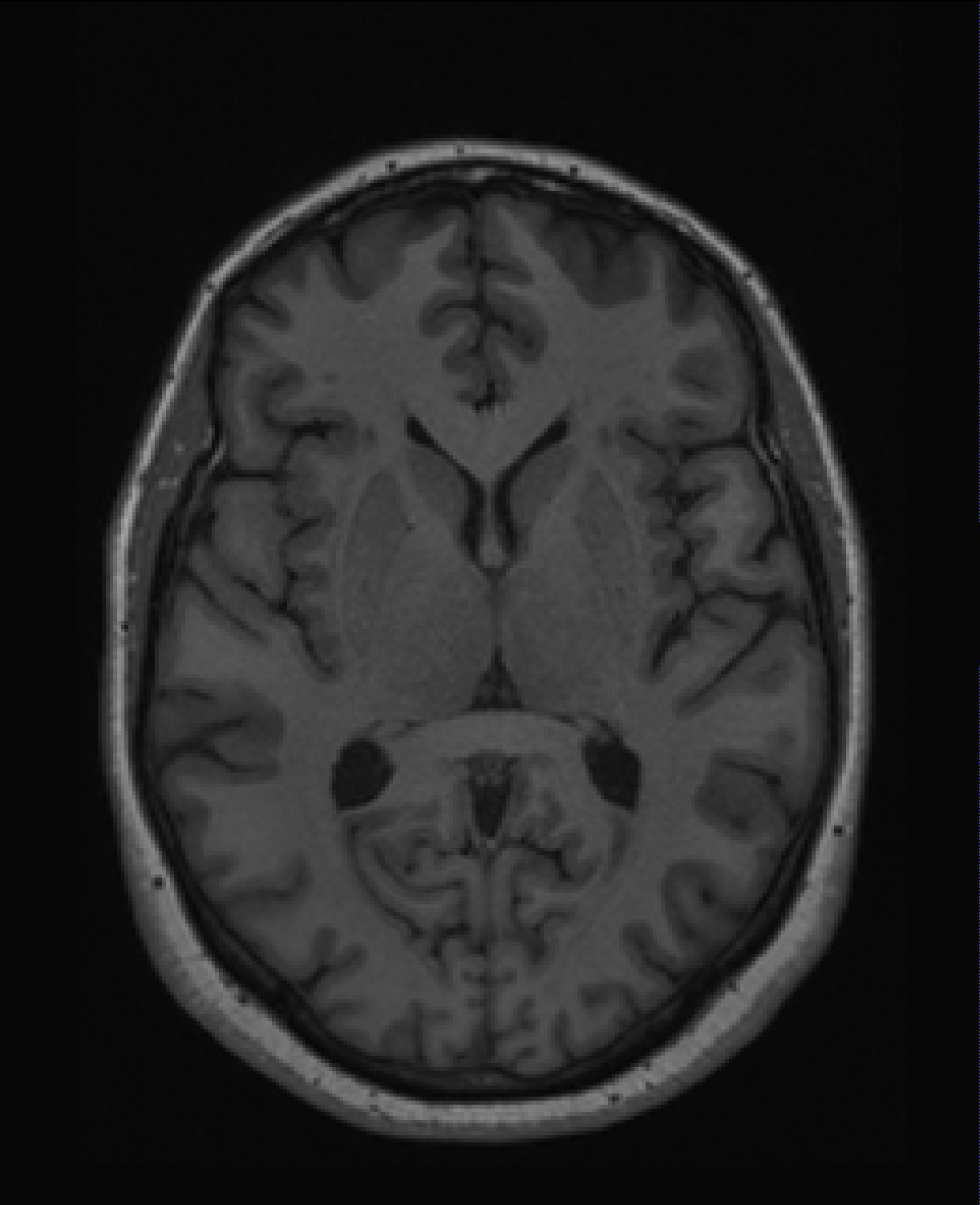}}
\hspace{0.5ex}
\subfloat[T2]{\includegraphics[width=0.17\textwidth]{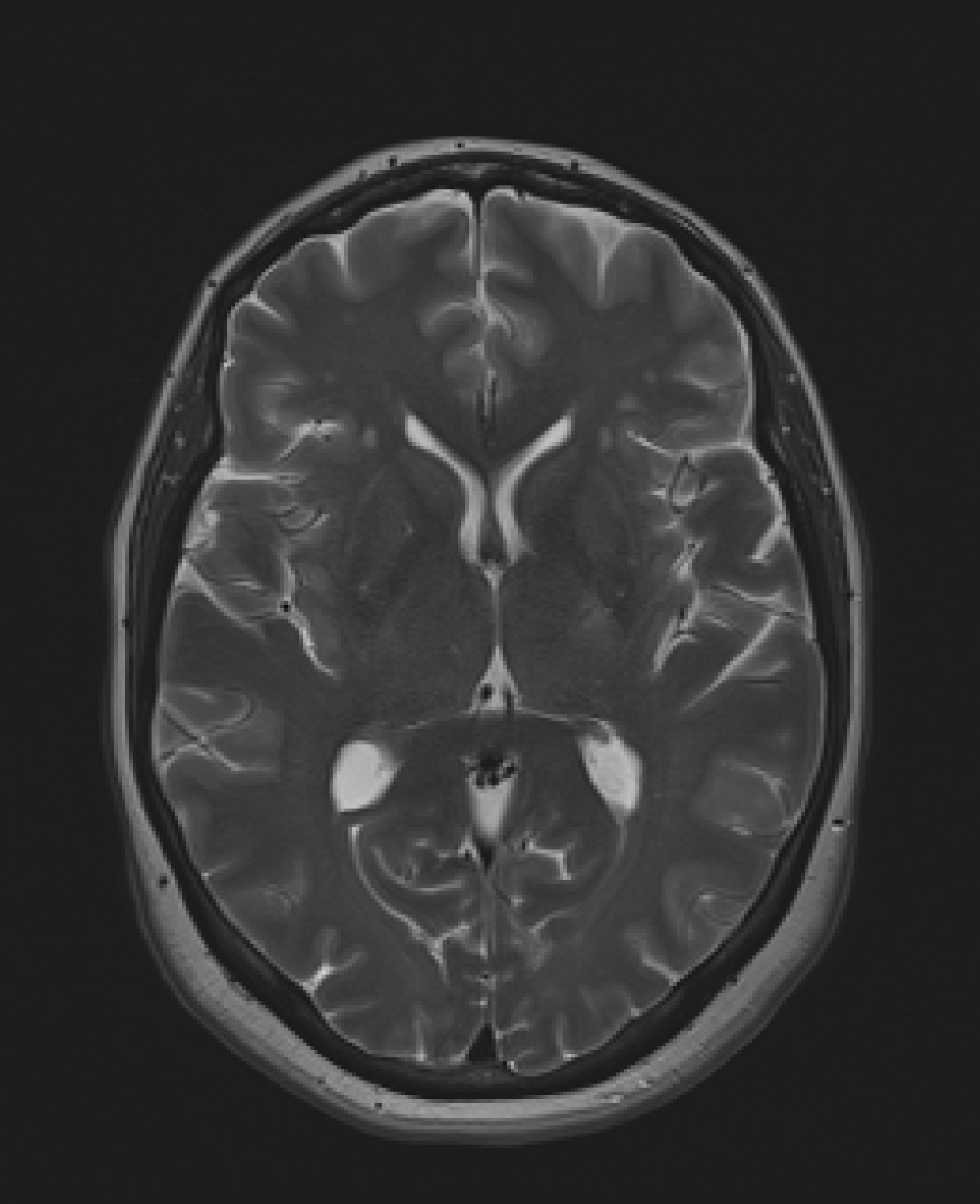}}
\hspace{0.5ex}
\subfloat[T2-FLAIR]{\includegraphics[width=0.17\textwidth]{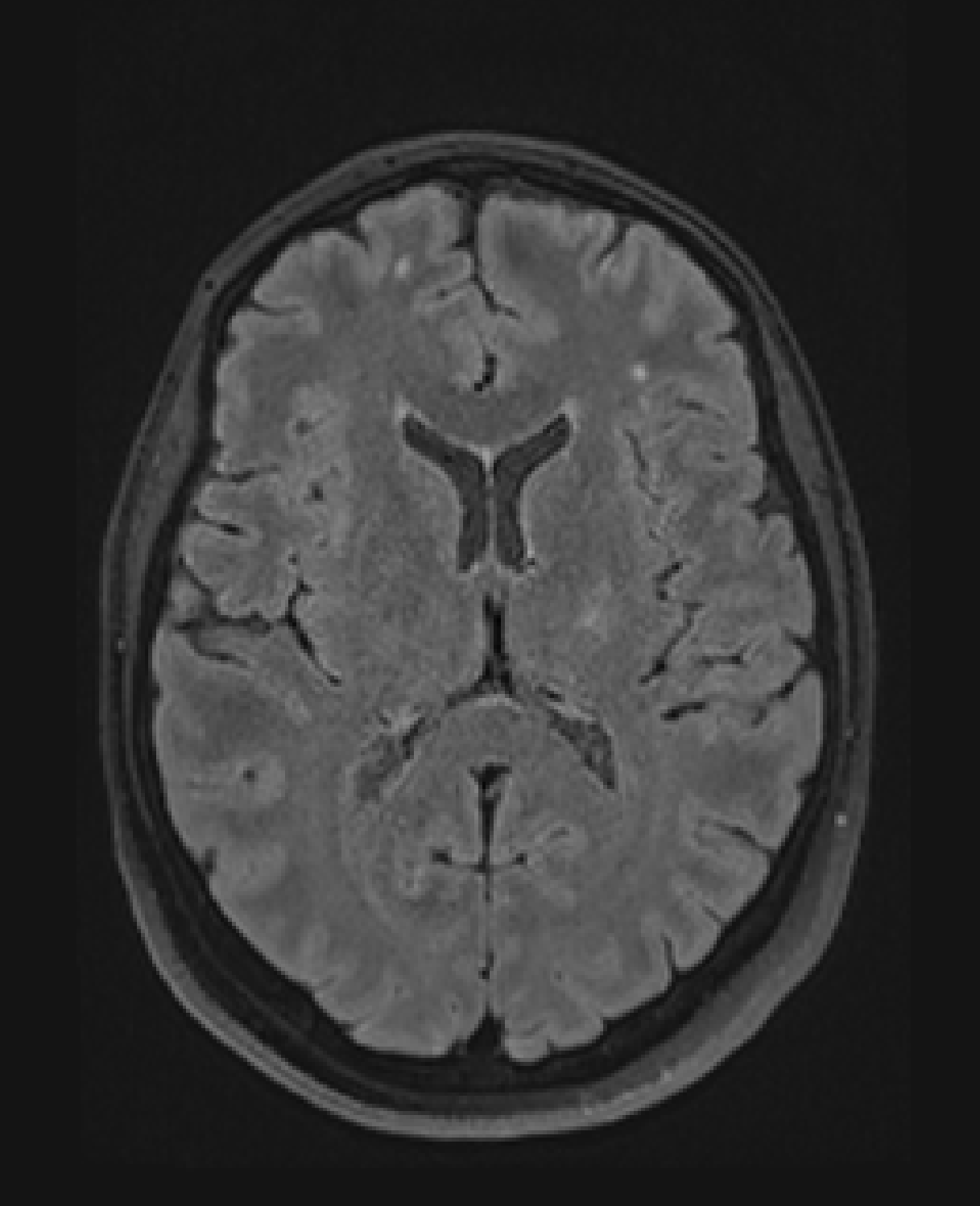}}
\hspace{0.5ex}
\subfloat[Golden Mask]{\includegraphics[width=0.17\textwidth,height=0.2088\textwidth]{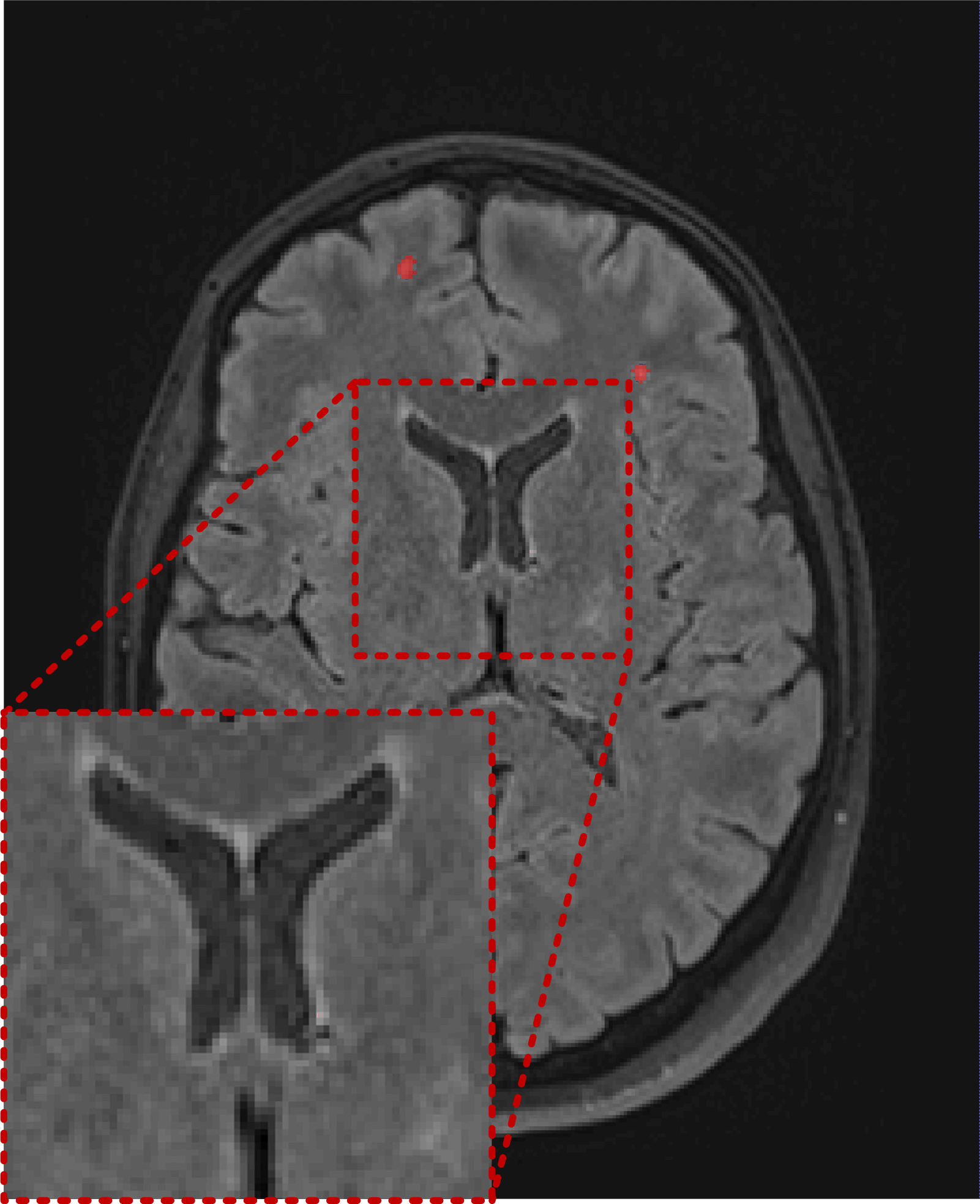}} \\
\subfloat[3D U-Net]{\includegraphics[width=0.17\textwidth,height=0.2088\textwidth]{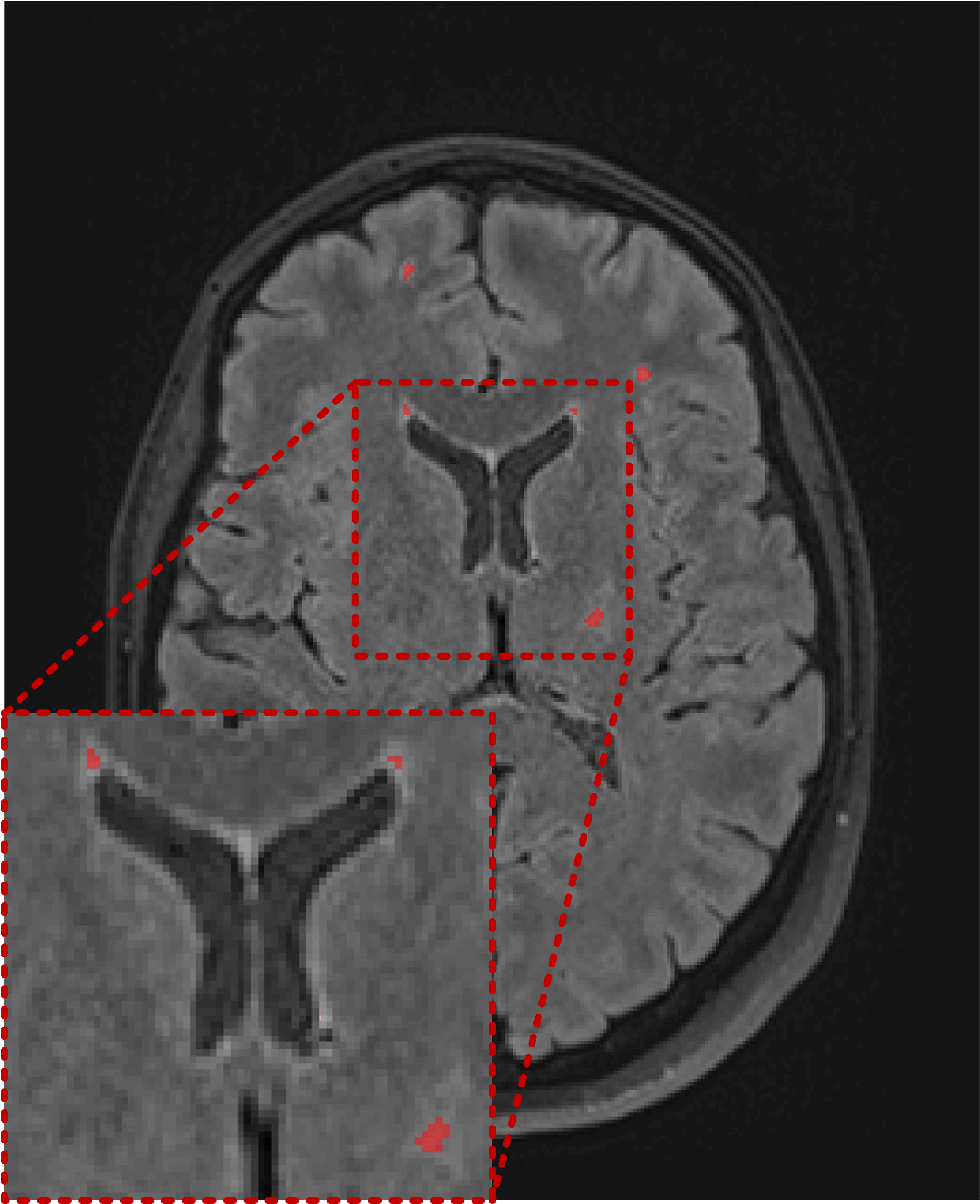}}
\hspace{0.5ex}
\subfloat[DA-Net]{\includegraphics[width=0.17\textwidth,height=0.2088\textwidth]{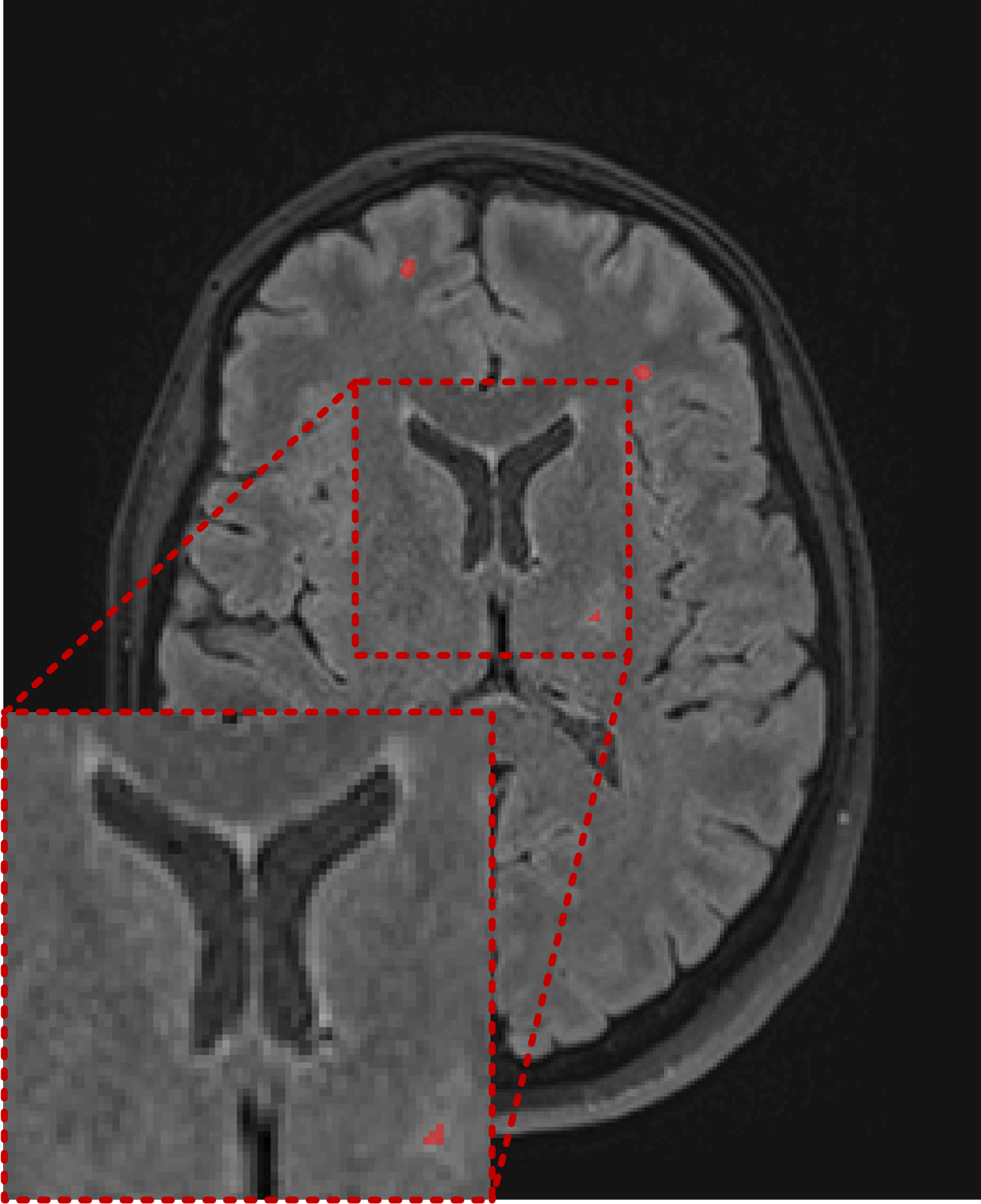}}
\hspace{0.5ex}
\subfloat[RSA-Net]{\includegraphics[width=0.17\textwidth,height=0.2088\textwidth]{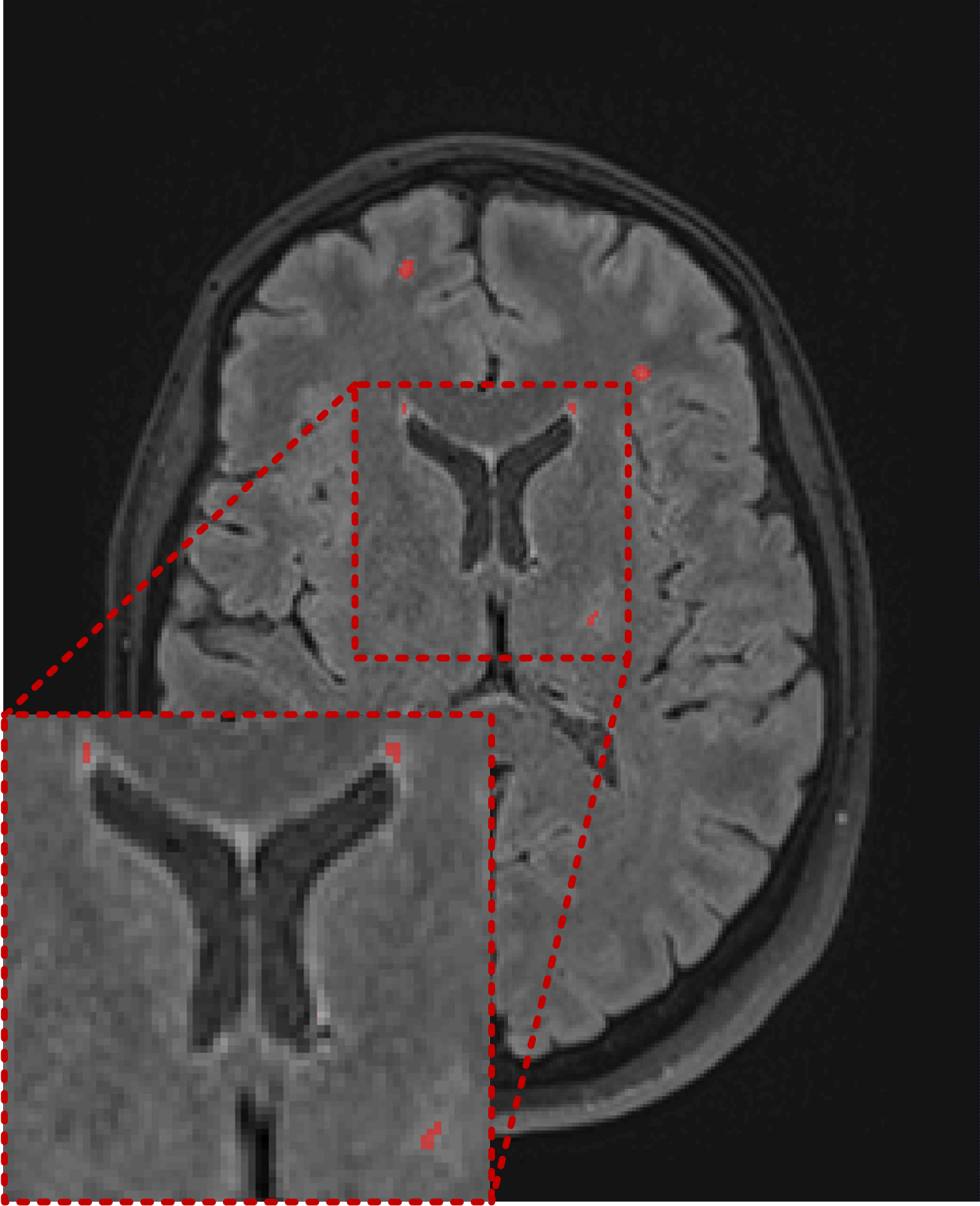}}
\hspace{0.5ex}
\subfloat[AG-Net]{\includegraphics[width=0.17\textwidth,height=0.2088\textwidth]{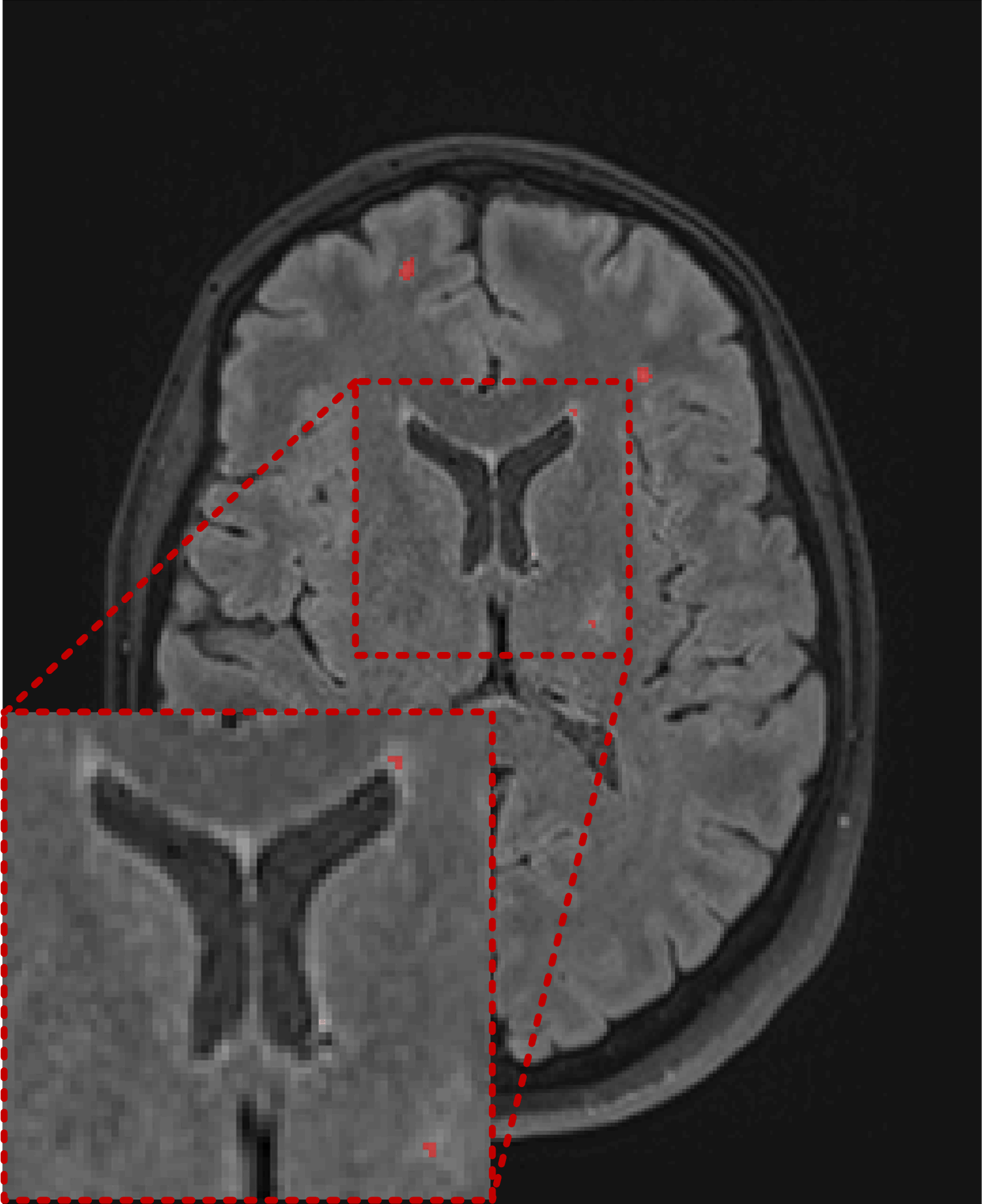}}
\hspace{0.5ex}
\subfloat[FA-Net]{\includegraphics[width=0.17\textwidth,height=0.2088\textwidth]{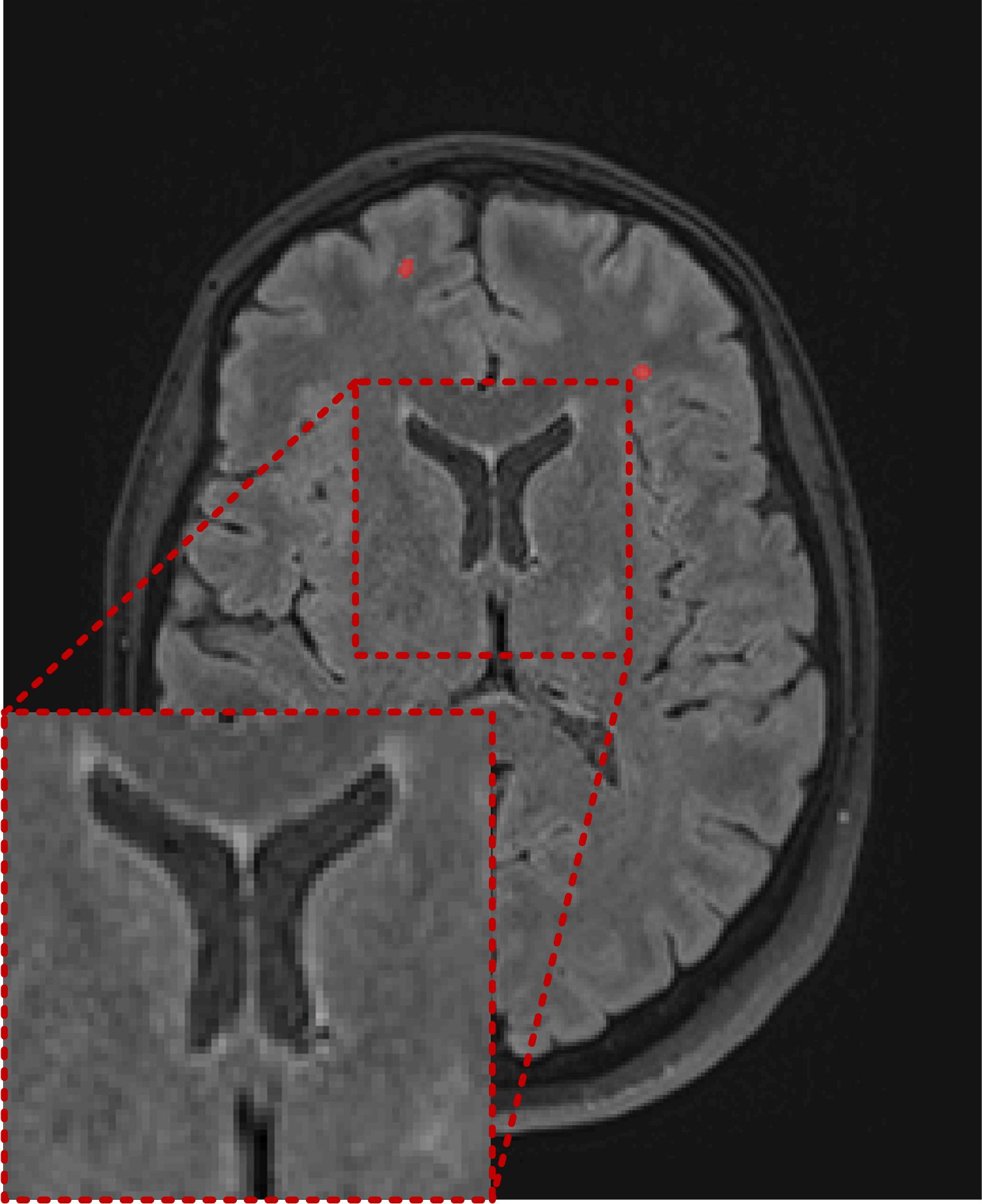}}
}
\caption{Example MS lesion segmentation results.
T1, T2, T2-FLAIR images and the corresponding golden mask are shown in the first line.
Segmentations from 3D U-Net, DA-Net, RSA-Net, AG-Net, and FA-Net are shown in the second line.
}
\label{fig:ms_lesion_results}
\vskip -0.1in
\end{figure*}

\subsubsection{Rank-One Constraint}

For any input feature tensor $\mathbf{X}$, we can compute four sub-affinity matrices $\mathbf{A}^{p_h} \in \mathbb{R}^{H\times H}, \mathbf{A}^{p_w} \in \mathbb{R}^{W\times W}, \mathbf{A}^{p_d} \in \mathbb{R}^{D\times D}$, and $\mathbf{A}^{p_c} \in \mathbb{R}^{C\times C}$.
Let $\mathbf{A}_i^{p}$ denotes the transpose of the $i_{th}$ row of the matrix $\mathbf{A}^{p}$.
We further define $\mathbf{A}_v \in \mathbb{R}^{H\times W\times D\times C}$ as follows:
\begin{equation}
    \mathbf{A}_v = \mathbf{A}_i^{p_h}\otimes \mathbf{A}_j^{p_w}\otimes \mathbf{A}_k^{p_d} \otimes \mathbf{A}_q^{p_c},
    \label{eq:tensor_product}
\end{equation}
where $v=(i,j,k,q)$ denotes the position of an element in the feature tensor and $\otimes$ is the tensor product; $\mathbf{A}_v$ is the affinity tensor of element $\mathbf{x}_v$ that shares the same size as input feature tensor, and all elements of the input feature tensor have their own affinity tensors.
The original affinity matrix $\hat{\mathbf{A}} \in \mathbb{R}^{NC\times NC}$ can be reconstructed using $\{\mathbf{A}_v | v \in \Omega \}$, where $ \Omega$ enumerates all possible element positions.
We can further derive $\mathbf{Z}_v$ as follows:
\begin{equation}
    \mathbf{Z}_v = \mathbf{A}_v \odot g(\mathbf{X})
    \label{eq:ele_sum}
\end{equation}
We have derived our full FA operation in Eq.~\ref{eq:full_fa}, 
It is easy to understand that enumerating all element positions using Eq.~\ref{eq:tensor_product} and Eq.~\ref{eq:ele_sum} can get the same result as using Eq.~\eqref{eq:full_fa}.
However, using Eq.~\ref{eq:full_fa} with a cascaded process can tremendously save the computational cost of GPU memory by taking advantages of replacing the original dense affinity matrix with four smaller sub-affinity matrices.
Since $\mathbf{A}_v$ is a rank-one tensor, FA can be considered as imposing an explicit low-rank constraint on the affinity tensor of each element.

\subsection{Complexity Analysis}

Given the input feature tensor of size $(H,W,D,C)$, we analyze the computational complexity of the proposed FA approach.
Let $N, M$ denotes the product and the sum of $H,W,D,C$, we can obtain the complexity of our FA as $\mathcal{O}(NC+NM)$.

We then numerically compare the computational complexity and GPU memory consumption of FA with other three approaches in Fig.~\ref{fig:memory_computation_overtime}. 
It can be seen that, comparing with DA~\cite{fu2019dual} that considers both spatial and channel attention, our FA is much more computational-efficient and GPU memory-friendly.
Though numerically, FA has little improvement of FLOPs over SA~\cite{wang2018non}, it reduces the GPU memory usage dramatically and in the meanwhile incorporates channel attention.
Comparing with FA, AG~\cite{oktay2018attention} has its limitation in 3D medical image applications as it still requires large memory for computation and suffers from scaling.

\section{Experimental Results}

\label{sec:experiments}

We use PyTorch~\cite{paszke2019pytorch} for all of our implementations.
We compare our models with several recent state-of-the-art attention approaches, including baseline 3D U-Net \cite{cciccek20163d}, dual attention (DA)~\cite{fu2019dual}, recurrent slice-wise attention (RSA)~\cite{zhang2019rsanet} and attention gated (AG) net~\cite{oktay2018attention}. 
For fairness, we adopt methods from their open-source implementations and do our best to adjust their parameters to achieve the best performance.
Particularly, DA is originally designed for 2D images, so it is modified and adjusted to be capable of processing 3D MR images.
All models in the experiments are trained in a machine with a Titan Xp GPU.

\subsection{Multiple Sclerosis (MS) Lesion Segmentation}

We conduct our first experiment on MS lesion segmentation, a high-level segmentation task. 
MS is a chronic, inflammatory demyelinating disease of central nervous system in the brain. 
Precise lesion tracing can provide important bio-markers for clinical diagnosis and disease progress assessment. 
However, MS lesion segmentation is challenging as lesions vary vastly in terms of location,
appearance, shape, and conspicuity (see Fig.~\ref{fig:ms_lesion_results} for more details). 

We use a dataset with 30 MR images acquired from a 3.0 T GE scanner. 
Images from T1, T2, and T2-FLAIR sequences are collected, and each voxel size is $0.7 \times 0.7 \times 3.0 mm^3$. 
Golden masks are traced by a neural radiologist with over 8 years’ lesion tracing experience.
Images are linearly co-registered using FLIRT at FSL~\cite{jenkinson2012fsl} neuroimaging toolbox.
All images are normalized to zero-mean with a unit-variance during the pre-processing step.


\begin{figure*}[!ht]
\centering 
\subfloat{\includegraphics[width=0.8\textwidth]{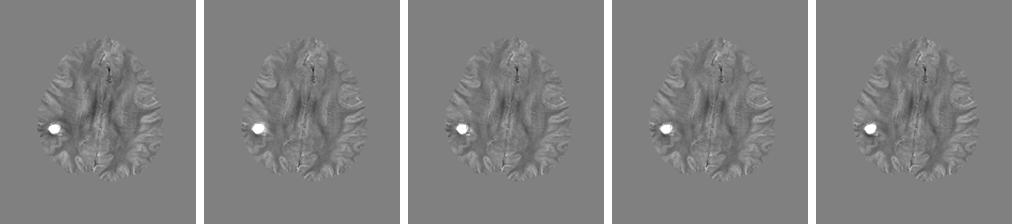}}
\vspace{1ex}
\subfloat{\includegraphics[width=0.8\textwidth]{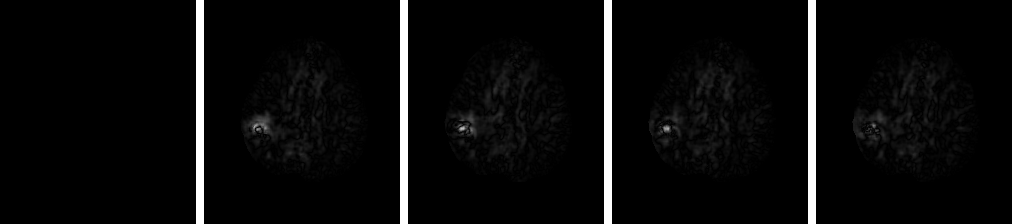}}
\caption{Example QSM reconstruction results (upper line with window level: [-0.15, 0.15] ppb) and absolute error maps (lower line with window level: [0, 0.05] ppb) on one test case with COSMOS label. 
From left to right are COSMOS (golden ground-truth), predictions of QSMnet, DA-Net, RSA-Net, and FA-Net.}
\label{fig:qsm1}
\vskip -0.1in
\end{figure*}

\subsubsection{Implementation Details}

We perform five random splits on the dataset, where each split contains 15, 5, and 10 subjects for training, validation, and testing.
A Model that achieves the minimum loss on the validation set will be used for testing.
We perform random crop with fixed cropping size ($128 \times 160 \times 32$), and use elastic deformation, intensity shifting for data augmentation.
We adopt the sum of weighted cross entropy and soft dice~\cite{dice1945measures} as our loss function.
Adam \cite{kingma2014adam} with the initial learning rate of $1\mathrm{e}{-3}$ and a multi-step learning rate scheduler with milestones at $50\%$, $70\%$ and $90\%$ of the total epochs are used for optimal convergence.
A batch size of four is used for training, and training would stop after $120$ epochs.

Dice score (DSC), lesion-wise true positive rate (LTPR), lesion-wise positive predicted value (LPPV), and lesion-wise F1 score (L-F1) are used for evaluations. 
LTPR and LPPV are defined as $\text{LTPR} = \dfrac{\text{TPR}}{\text{GL}},\text{LPPV} = \dfrac{\text{TPR}}{\text{PL}}$,
where TPR denotes the number of lesions in the Golden segmentation that overlaps with a lesion in the produced segmentation, and GL, PL is the number of lesions in ground-truth segmentation and produced segmentation respectively.
$\text{L-F1}$ can be obtained from $\text{LTPR}$ and $\text{LPPV}$ as $\text{L-F1} = 2 \dfrac{\text{LTPR}\cdot \text{LPPV}}{\text{LTPR}+\text{LPPV}}$.

\subsubsection{Quantitative Results}

We use DA-Net, RSA-Net, and FA-Net to denote a backbone 3D U-Net with the corresponding attention module inserted at the bottom layer of 3D U-Net.
Specifically, AG-Net inserts three attention modules according to the literature \cite{oktay2018attention}.
As shown in Table~\ref{ms_lesion_2}, all attention methods outperform 3D U-Net backbone network in all metrics by a significant margin.
RSA-Net and AG-Net have no clues about dependencies or salience of channels; thus, we can see from the table that our FA-Net outperform them in both DSC and L-F1 metrics; Though RSA-Net obtains similar LTPR as our FA-Net, it falls behind a lot in LPPV.
Though DA-Net considers both spatial and channel attention and our FA-Net has only marginal improvement compared to DA-Net, incorporation of our FA module consumes negligible additional GPU memory and FLOPs (See Fig.~\ref{fig:memory_computation} and Fig.~\ref{fig:memory_computation_overtime}).
The superiority of FA-Net is that it considers feature aggregation from all elements of an input tensor by replacing original affinity matrix into several sub-affinity matrices and aggregate features in a cascading manner.

\begin{table}[!bh]
\vskip -0.1in
\caption{Quantitative comparison of MS lesion segmentation with different approaches.}
\label{ms_lesion_2}
\begin{center}
    \resizebox{1.02\columnwidth}{!}
    {
    \begin{small}
    \begin{tabular}{l|cccc}
    \hline
    \hline
    Method                           & DSC   & LPPV  & LTPR  & L-F1  \\
    \hline
    3D U-Net \cite{cciccek20163d}    & 0.667      & 0.682      & 0.838      & 0.752 \\
    DA-Net \cite{fu2019dual}         & 0.682      & 0.689      & \bf{0.871} & 0.770 \\
    RSA-Net \cite{zhang2019rsanet}   & 0.677      & 0.678      & 0.870      & 0.762 \\
    AG-Net \cite{oktay2018attention} & 0.682      & 0.702      & 0.830      & 0.761 \\
    FA-Net (ours)                    & \bf{0.684} & \bf{0.703} & 0.867      & \bf{0.776} \\
    \hline
    \hline
    \end{tabular}
    \end{small}
}
\end{center}
\vskip -0.1in
\end{table}

\subsubsection{Qualitative Results}

We showcase one slice from a testing subject, and compare the qualitative results of different models with the golden mask.
We can see from Fig.~\ref{fig:ms_lesion_results} that besides MS lesions, there still exists many other concurrent hyper-intensities in the T2-FLAIR image.
Particularly, the hyper-intensities near the lateral ventricles are prone to be over-segmented. 
This is because some hyper-intensities near ventricles are MS lesions, but some are not, depending on their anatomical and surrounding structures.
We can see that all attention models help ease the over-segmenting problem in some degree. 
DA-Net and our FA-Net perform the best as these two models both consider the dependencies of spatial and channel dimensions.

\begin{figure*}[!ht]
\centering {
}
\subfloat[QSMNet]{\includegraphics[width=0.17\textwidth]{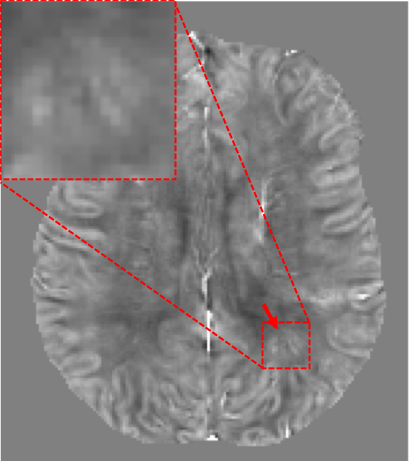}} \hspace{1ex}
\subfloat[DA-Net]{\includegraphics[width=0.17\textwidth]{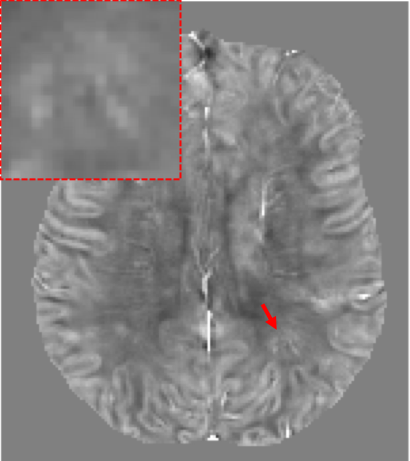}}  \hspace{1ex}
\subfloat[RSA-Net]{\includegraphics[width=0.17\textwidth]{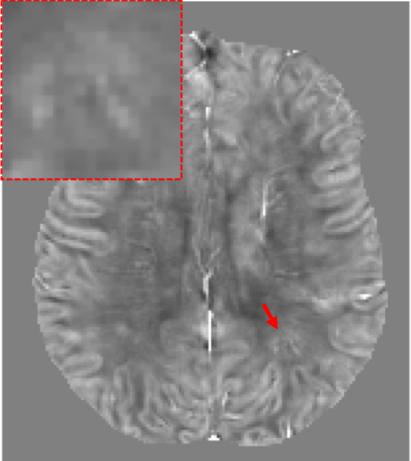}} \hspace{1ex}
\subfloat[FA-Net]{\includegraphics[width=0.17\textwidth]{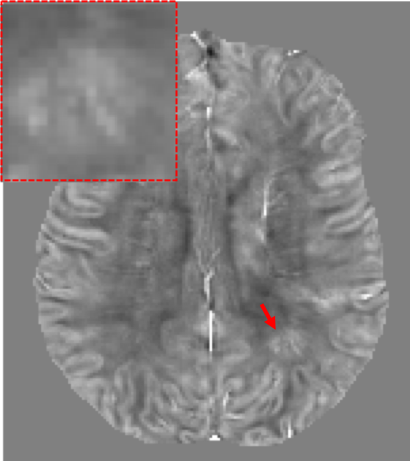}}  \hspace{1ex}

\caption{Example QSM reconstruction results (window level: [-0.15, 0.15] ppb) on a subject with MS lesions (patient data without ground-truth COSMOS). 
Hyperintense MS lesions are pointed out by red arrows.
}
\label{fig:qsm2}
\vskip -0.1in
\end{figure*}

\subsection{Quantitative Susceptibility Mapping (QSM)}
We conduct our second experiment on a challenging image reconstruction problem in MRI: quantitative susceptibility mapping (QSM) \cite{de2010quantitative, wang2015quantitative}. 
QSM is a recently developed image contrast that can measure the underlying tissue apparent magnetic susceptibility, which can be used to quantify specific bio-markers such as iron, calcium, and gadolinium. 
The forward model of generating magnetic field from susceptibility map with additive noise is a three-dimensional spatial convolutional process and can be described as following:
\begin{equation}
    b = \chi * d + n,
    \label{eqn:qsm}
\end{equation}
where $b$ is the magnetic field, $\chi$ is the tissue susceptibility, $d$ is the dipole convolution kernel, and $n$ is the additive measurement noise. 
The aim of QSM is to solve the deconvolutional problem from measured noisy magnetic field $b$ to tissue susceptibility $\chi$. 
This is intrinsically an ill-posed inverse problem due to the zero cone surfaces of the dipole kernel in k-space \cite{wang2015quantitative}. 
To tackle the ill-poseness, COSMOS (Calculation Of Susceptibility through Multiple Orientation Sampling) \cite{liu2009calculation} reconstruction is proposed to eliminate all zeros in the k-space cone surface by multiple orientation scans, thereby serving as the golden susceptibility for further clinical analysis.

Recently, several deep learning based QSM reconstruction methods~\cite{yoon2018quantitative, zhang2020fidelity, zhang2020bayesian, chen2020qsmgan} have been developed with promising results. 
They use 3D U-Net as the backbone network to perform the functional mapping from the magnetic field input to susceptibility output. 
In this experiment, we follow previous work and use COSMOS data to train our deep networks. 
To acquire and reconstruct COSMOS data, $6$ healthy subjects were recruited to do MRI scan with $5$ brain orientations using a 3.0T GE scanner (Please note that COSMOS technique cannot be applied to patients as it needs five times as much time to scan each patient). 
Acquisition matrix was $256\times256\times48$ and voxel size was $1\times1\times3 \ \text{mm}^3$. 
Golden tissue susceptibility is reconstructed with five orientations of each subject using COSMOS, and local magnetic field data is generated accordingly using Eq.~\eqref{eqn:qsm}.

\begin{table}[!hb]
\vskip -0.1in
\caption{Quantitative comparison of QSM.}
\label{table:qsm1}
\begin{center}
    \resizebox{1.02\columnwidth}{!}
    {
        \begin{small}
        \begin{tabular}{l|cccc}
        \hline
        \hline
        Method & RMSE & HFEN & SSIM & PSNR \\
        \hline
        QSMnet \cite{yoon2018quantitative} & $31.99$ & $33.37$ & $0.9824$ & $48.86$ \\
        DA-Net \cite{fu2019dual}           & $32.15$ & $33.84$ & $0.9826$ & $48.78$ \\
        RSA-Net \cite{zhang2019rsanet}     & $31.65$ & $33.18$ & $0.9830$ & $48.91$ \\
        FA-Net (Ours)                      & $\bf{31.18}$ & $\bf{32.49}$  & $\bf{0.9833}$ & $\bf{49.06}$ \\
        \hline
        \hline
        \end{tabular}
        \end{small}
    }
\end{center}
\vskip -0.1in
\end{table}

\subsubsection{Implementation Details}

We perform six splits on the dataset, where each split contains 4, 1, and 1 subjects(s) for training, validation, and testing, and each subject contains 5 volumes. 
During training, we cropped each volume into 3D patches in size ($64\times 64\times 32$) and use in-plane rotation of $\pm 15^\circ$ for data augmentation. 
Loss function from QSMnet~\cite{yoon2018quantitative} is adopted. 
Adam \cite{kingma2014adam} optimizer is used for training with the same hyper-parameters as MS lesion segmentation experiment. 
Training is performed with a batch size of $16$ and training would stop after $60$ epochs .
During testing, a model with the best validation loss is used to evaluate the performance. 
In addition, a patient subject with MS lesion is also used to qualitatively verify the performance of our networks. (Note that a patient subject does not have the COSMOS ground-truth)
Different from MS lesion segmentation, we use QSM-Net~\cite{yoon2018quantitative}, a modified U-Net, as our backbone network.
We use DA-Net, RSA-Net, and FA-Net to denote a QSM-Net with the corresponding attention module inserted at its bottom layer.
AG-Net is excluded in the QSM experiment as it is unfair to compare MA based methods with SA based methods in a full functional image mapping task.

\paragraph{Quantitative Results}

We use root mean square error (RMSE), peak signal-to-noise ratio (PSNR) (measures general reconstruction error), high-frequency error norm (HFEN) (measures the similarity at high spatial frequencies), and structural similarity index (SSIM) (quantifies image contrast, intensity, structural similarity between image pairs \cite{wang2004image}) to quantify the reconstruction accuracy.
Quantitative results averaged among six splits are shown in Table ~\ref{table:qsm1}, and we can see that our FA-Net shows the best reconstruction results in all four metrics. 

\paragraph{Qualitative Results}

We choose one slice from the testing image of one split, and the chosen subject is diagnosed as cerebral hemorrhage (hyper-intensity tissue area in Fig.~\ref{fig:qsm1}); however, the hemorrhage situation is not covered in the training data. 
As we can see from Fig.~\ref{fig:qsm1}, the error map from our FA-Net achieves the minimum intensity which shows the robustness of our FA-Net compared to others.

We use an additional MS lesion subject without ground-truth COSMOS to compare the reconstruction performance among four trained networks in Fig.~\ref{fig:qsm2}. 
As can be seen from Fig.~\ref{fig:qsm2}, on one hand, our FA-Net generated the most hyperintense lesions, and on the other hand, the lesion shows clearer boundary in FA-Net produced image compared to others.
The superiority of our FA-Net is that it aggregates features from both spatial and channel dimensions, and in the meanwhile, it regularizes the dense affinity matrix with rank-one constraint and thus generalizes better to unseen situations.
\section{Conclusions}

\label{sec:conclusions}

We presented a novel folded attention module. 
Our FA module exploits the spatial-channel correlations in an efficient and effective way.
FA not only achieves the highest accuracy on MS lesion segmentation and QSM reconstruction among all state-of-the-art attention methods, but also reduces tremendously the computational overhead and memory usage.
Our method can be easily plugged into any existing CNN model with negligible cost, thereby serving as a new baseline for general 3D MR image processing.

\newpage

\newpage


\bibliographystyle{aaai}
\bibliography{egbib.bib}

\end{document}